\documentclass[10pt,conference]{IEEEtran}
\usepackage{amsmath,amsthm,amsfonts,amssymb,bm,mathrsfs}
\usepackage{graphicx,subfigure}
\usepackage[bookmarks=false,colorlinks,citecolor=black, filecolor=black, linkcolor=black,urlcolor=black]{hyperref}
\usepackage{url}
\usepackage{algorithm} 
\usepackage{algorithmic}
\ifCLASSINFOpdf
\else
\fi
\usepackage{multirow}
\usepackage{graphicx}
\usepackage{epstopdf}
\usepackage{epsfig}
\begin{document}
%
\title{AI-based Two-Stage Intrusion Detection for Software Defined IoT Networks}


\author{\IEEEauthorblockN{Jiaqi Li, Zhifeng Zhao, Rongpeng Li and Honggang Zhang}
\IEEEauthorblockA{College of Information Science $\&$ Electronic Engineering, Zhejiang University\\
Email: \{21631097, zhaozf, lirongpeng, honggangzhang\}@zju.edu.cn}}
\thanks{J. Li, Z. Zhao, R. Li, and H. Zhang are with College of Information Science and Electronic Engineering, Zhejiang University. Email: \{21631097, lirongpeng, zhaozf, honggangzhang\}@zju.edu.cn}
\thanks{This work was supported in part by the Program for Zhejiang Leading Team of Science and Technology Innovation (No. 2013TD20), in part by the National Postdoctoral Program for Innovative Talents of China (No. BX201600133), and in part by the Project funded by China Postdoctoral Science Foundation (No. 2017M610369).}


%


\maketitle

\begin{abstract}
Software Defined Internet of Things (SD-IoT) Networks profits from centralized management and interactive resource sharing which enhances the efficiency and scalability of IoT applications. But with the rapid growth in services and applications, it is vulnerable to possible attacks and faces severe security challenges. Intrusion detection has been widely used to ensure network security, but classical detection means are usually signature-based or explicit-behavior-based and fail to detect unknown attacks intelligently, which are hard to satisfy the requirements of SD-IoT Networks. In this paper, we propose an AI-based two-stage intrusion detection empowered by software defined technology. It flexibly captures network flows with a globle view and detects attacks intelligently through applying AI algorithms. We firstly leverage Bat algorithm with swarm division and Differential Mutation to select typical features. Then, we exploit Random forest through adaptively altering the weights of samples using weighted voting mechanism to classify flows. Evaluation results prove that the modified intelligent algorithms select more important features and achieve superior performance in flow classification. It is also verified that intelligent intrusion detection shows better accuracy with lower overhead comparied with existing solutions. 
\end{abstract}


\section{Introduction}
Internet of Things (IoT) is an evolving technology which provides ubiquitous connectivity and interaction between the physical and cyber worlds \cite{Al2015Internet}. However, the rapid increase in the number and diversity of smart devices connected to the Internet has raised the issues of flexibility, efficiency and availability within the current IoT networks. As an essential trend of Iot networks, the emergence of Software Defined IoT networks \cite{Mahdavinejad2017Machine} provides a manageable solution which has drawn significant attention. Benifiting from the advantages of Software Defined Network (SDN) \cite{Kim2013Improving}, SDN-based approach facilitates the supervision of network status and the collection of information under centralized control in an active manner. Moreover, it also optimizes network management and resource allocation flexibly through software programmability to meet the diverse demands of IoT networks. Nevertheless, the development of SD-IoT dose not totally eliminate various security issues and challenges. In order to address these emerging problems, it is urgent to build up effective and intelligent algorithms for enforcing the security of Software Defined IoT networks \cite{Yassein2018Combined}.

As an indispensable technology in network security, intrusion detection mechanisms dynamically monitor the abnormal behaviors or patterns in a system and indicate whether some events are susceptible of an attack \cite{Tang2016Deep}. There are two main categories of intrusion detection techniques: misuse detection and anomaly detection. Misuse detection are usually signature-based, which can only detect known attacks by matching the behaviors of incoming intrusions with the historical knowledge and predefined rules. Anomaly detection automatically constructs a normal behavior of the systems and stubbornly detects incoming intrusions by explicitly computing deviations. It can recognize novel attacks but may raise false alarms as well. 

To overcome the limitation of traditional intrusion detection, Artificial Intelligence (AI) has been taken into acount for intelligent detection. AI-based schemes can automatically discover deep knowledge or patterns from the historical data and make wise judgments to predict network intrusions \cite{Buczak2017A}\cite{Li2017Intelligent}. Though, there have been a few researches on combining IDS and AI, they are still incapable for universally precise detection and robustly considering the evolution and development of SD-IoT networks.

In this paper, we propose an advanced intrusion detection technology using AI algorithms based on Software Defined IoT architecture. Specifically we apply a combination of enhanced AI algorithms to perform feature selection and flow classification, which are two crucial steps in intrusion detection. In particular for feature selection, we take advantage of improved Bat algorithm by splitting the whole swarm into subgroups using K-means method so that each subgroup can learn within and among different populations more efficiently. Besides, Differential Evolution is also employed to increase the diversity of individuals. For flow classification, we optimize Random forest through updating the weight of each sample after building each tree iteratively and making the final decision by using weighted voting mechanism.

The remainder of the paper is organized as follows. Section \ref{sec:related work} discusses the related research works. Section \ref{sec:overview} introduces the AI-based two-stage intrusion detection  empowered by Software Defined technology. Section \ref{sec:algorithm1} describes the proposed BA algorithm for feature selection. Section \ref{sec:algorithm2} presents the improved RF algorithm for traffic classification. Section \ref{sec:evaluation} presents the performance evaluation results. Section \ref{sec:conclusion} summarizes the paper and points out the potential future work.

\section{Related work}
\label{sec:related work}
As a critical enabling technology, SDN radically revolutionizes the way network operators will architect and coordinate, and meets the demands of IoT networks in terms of performance and reliability. It dynamically manages network configurations and provides flexible service provisioning mechanisms in a centralized control manner \cite{Zhang2016Cloudified}. Accordingly, the combination of IoT networks and SDN  has attracted tremendous research interests. In \cite{Ojo2017A}, an architecture with SDN based IoT framework coupled with network functionality virtualization (NFV) is introduced. It provides a general implementation by virtualizing the IoT gateway which makes it possible to be dynamic, scalable and elastic in the IoT networks. Another IoT architecture originated from SDN to overcome big data problem is proposed in \cite{Kakiz2017A}. By evaluating the usefulness of the sensed values in the lower layers (especially in gateway layer) instead of application layer, the number of packets being sent to the Internet is reduced, which overcomes the huge data volume problem of IoT. 

The advancement of SDN has strengthened IoT security through supporting the supervision of network status and the collection of flows statistics. It also provides network-layer security services such as packet routing, identity authentication, and automated security management in a global view, which facilitates the detection and prevention of attacks \cite{Rawat2017Software}. However, SD-IoT networks still face severe security challengesa as new attacks quikly appear. Therefore, several previous studies have investigated the ability of SDN and introduces various solutions to improve IoT security. In \cite{Salman2016Identity}, it proposes an identity-based authentication scheme for IoT based on SDN. The specific identity formats used by different communication protocols are mapped to a shared identity and a trusted certificate authority is implemented on the SDN controller. \cite{Nobakht2016A} also proposes a host-based intrusion detection and mitigation framework for IoT, in which network visibility and flexibility properties of SDN are exploited and modules of intrusion detection and mitigation are implemented at the SDN controller. Moreover, remote security management is provided by the third-party entities that offer `Security as a Service'. A flow-based security approach for IoT devices is proposed using an SDN gateway in \cite{Bull2016Flow}. It aims to mitigate Distributed Denial of Service (DDoS) attacks that violates services availability by monitoring traffic flows and anomalous behavior. These proposed schemes are applicable to solve specific types of security problems and can only perform well under certain scenarios.

\begin{figure}[!t]
	\centering{\includegraphics[scale=0.3]{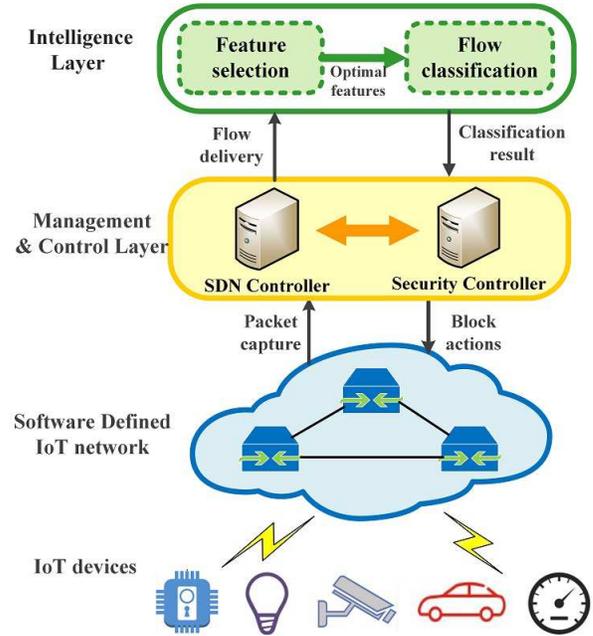}}
	\caption{AI-based Two-Stage Intrusion Detection for
		Software Defined IoT Networks}
\end{figure}

As the focus of network security, intrusion detection has gained intensive attentions and wide investigations in recent years \cite{Farah2015Application}. Most of them depend on pre-defined rules, which are still unable to recognize uprising new attacks intelligently. In terms of this issue, various machine learning algorithms have been adopted with flow-based classification in solving such problems \cite{Silva2016ATLANTIC}. Most often, there are two stages in this process: feature selection and flow classification. The former stage addresses high dimension data efficiently and decreases the number of features from a noisy dataset, which improves the learning efficiency and prediction accuracy of flow classification. Recently, a variety of Swarm Intelligence (SI) algorithms such as ant colony optimization (ACO) \cite{Mehmood2016SVM} and particle swarm optimization (PSO) \cite{Cleetus2014Multi} have been applied to select optimal features as well. The latter stage distinguishes network flows by marking whether it belongs to specific types of attacks or benign traffic. \cite{Le2016Flexible} proposes a network-based IDPS (Intrusion Detection and Prevention System) which performs C4.5 algorithm to build the decision tree for classifying the traffic. In \cite{Kim2016Long}, the authors propose a three-layer Recurrent Neural Network (RNN) which is capable of automatically finding the correlations between flow records and acting as a neural classifier for misuse detection. However current algorithms are still inadequate for effectively selecting optimal features and detecting new types of attacks with lower cost under different circumstances owing to their inherent limits. Therefore, we need to optimize existing algorithms for the two critical stages to detect network intrusions efficiently and adaptively.

\section{AI-based two stage intrusion detection}
\label{sec:overview}
In this paper, we propose a two-stage intrusion detection using AI algorithms under SD-IoT Networks, see Figure 1. Taking advantage of SDN, it captures network packets and collects status information with centralized control. The Controllers partitions network packets into flows and delivers them to the upper layer. In this way, flow-based intelligent intrusion detection can be implemented using AI algorithms in two stages. It selects optimal flow features and detects network anomalies by classifying each flow into specific catogories. Afterwards, the Controllers manages resource arrangement and organizes specific actions for defending attacks according to the classification results. 

We combine two artificial intelligent algorithms in the two main stage of intrusion detection. Swarm Intelligence (SI) algorithms have been widely adopted for global optimization through heuristic searching iteratively combined by classification which contributes to higher accuracy.
As a promising novel SI algorithm, Bat Algorithm (BA) solves feature selection problems and behaves better than traditional SI algorithms with simple structure, fewer parameters and stronger robustness. It achieve outstanding performance owing to its flexibility, simplicity, and robustness. Therefore in the fist stage, we enhance BA to improve its ability for searching optimal features. As an ensemble method, Random forest (RF) can prevent overfitting and make final dicision using majority voting. It has been validated that it outperforms other algorithms in various situations in terms of prediction accuracy with tolerable time complexity, which gains much more popularities in classification. So in the second stage, we optimize RF algorithm to classify the network traffic into different classes of attacks with the selected features as input. 

\section{Improved of Bat Algorithm for feature selection}
\label{sec:algorithm1}

Bat algorithm \cite{Enache2015A} is a metaheuristic algorithm for global optimization. It was inspired by the echolocation behaviour of microbats, with varying pulse rates of emission and loudness, which achieves great performance in But it still suffers from getting trapped into local minima easily since the position of each bat is only strongly influenced by the global best individual without communicating with its neighbors, which lacks diverse positions of the swarm. Also, the algorithm lacks of a mutation mechanism, which can not escape from local minima once the bat is adjacent to it. Thus, in order to address these problems, we enhance the algorithm as in the following two ways.

\begin{figure}[!t]
	\centering{\includegraphics[scale=0.045]{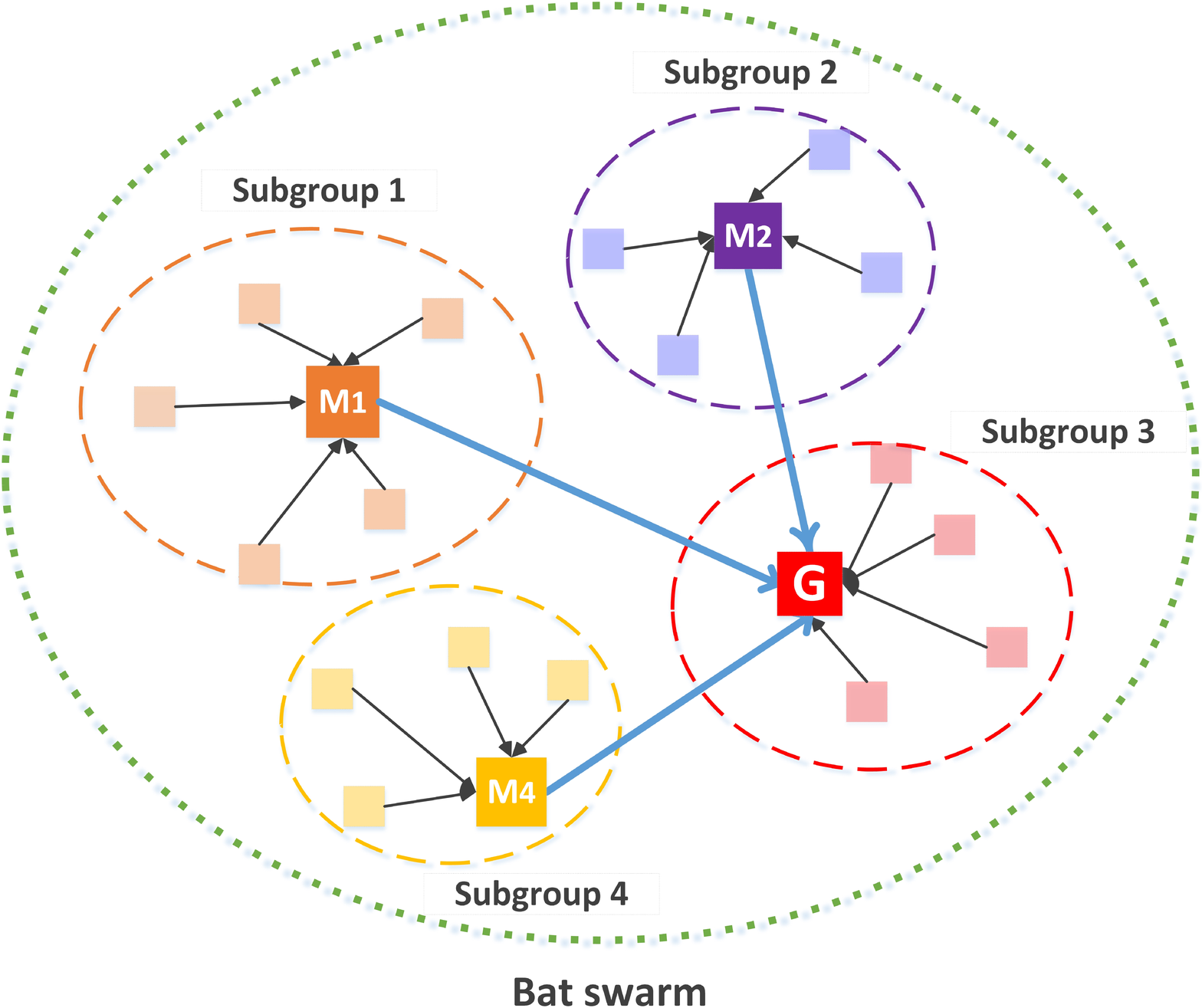}}
	\caption{Swarm division}
\end{figure}

\subsection{Swarm division}
At each iteration during the process, we divide the whole swarm (the total number of individuals is $N$) into several subgroups (e.g. $K$) containing the same number of individuals using K-means algorithm referring to the distances between them, see Figure 2. For each subgroup $n$ ($n=1,2,...K$), we select the local minima whose position is $M_{n}$. For all the subgroups, the global best bat is found and locates in $G$. Each bat retains its previous best position as $P_{i}$ ($i=1,2,...N$). For the $i$th individual, the flight is described by its location in space $x_i = (x_{i,1},x_{i,2} ,...,x_{i,d} )$ and velocity $v_i = (v_{i,1},v_{i,2},...,v_{i,d})$,
where $d$ is the problem dimension. $f_{i}$ is the frequency of the bat. The bat updates its velocity at iteration $t$ in two typical situations as follows:

For the non-local-minima individuals in each subgroup

\begin{small}
\begin{equation}
\label{eq:flow profile}
\begin{aligned}
v_{ij}^{t}= W^t\cdot v_{ij}^{t-1}+(x_{ij}^{t-1}-M_{n}^{t-1})\cdot f_{i}+(x_{ij}^{t-1}-P_{i}^{t-1})\cdot C^t
\end{aligned}
\end{equation}
\end{small}

For the local-minima individuals of each subgroup

\begin{small}
\begin{equation}
\label{eq:entropye}
\begin{aligned}
v_{ij}^{t}= W^t\cdot v_{ij}^{t-1}+(x_{ij}^{t-1}-G^{t-1})\cdot f_{i}+(x_{ij}^{t-1}-P_{i}^{t-1})\cdot C^t
\end{aligned}
\end{equation}
\end{small} 

In each iteration, $W^t$ is the inertia weight for each bat and $C^t$ is the self-learning factor of each bat. $W_{max}$, $W_{min}$, $C_{max}$, $C_{min}$ are the maximum and the minimum of $W$ and $C$, respectively. $N_{T}$ is the the largest number of iterations. The two variables are calculated as below.

\begin{small}
\begin{equation}
\label{eq:entropye}
\begin{aligned}
&W^t= W_{max} - \frac{(W_{max}-W_{min})\cdot t}{N_{t}}
\end{aligned}
\end{equation}
\end{small}
\begin{small}
\begin{equation}
\label{eq:entropye}
\begin{aligned}
&C^t= C_{min} +(C_{max}-C_{min}) \cdot(1-\frac{\arccos[(\frac{(-2)\cdot t}{N_{t}})+1] }{\pi})
\end{aligned}
\end{equation}
\end{small}

We introduce the linearly decreasing inertia weight $W^t$ varying as the iterations increase to improve the optimization ability of the algorithm. At early iterations, the bat with a larger inertia weight as well as a great speed is enabled with a strong global search ability. Later, a smaller inertia weight contributes to a more accurate local search, which accelerates the rate of convergence.

By employing the parameter $C^t$, we slightly enhance the velocity update  by taking advantage of the historical experience of the bat. The dynamically adjusting parameter indicates the influence of the previous best position of each bat on the current speed. The bat with larger $C^t$ retains its own position with a better exploration ability initially.  Afterwards, the position of the bat tends to be greatly affected by the global best bat which elevates its exploitation ability.

\subsection{Binary Differential Mutation}

After the update of a bat at each iteration, we further apply the mutation mechanism of Differential Evolution \cite{Wang2017Differential} to BA algorithm, which enhances the diversity of population and the ability of bats to jump out of local optimum. It disturbs the target value by using the differences of random selected individuals in the swarm. Since the original method can only solve continuous optimization problem, we put forward a Binary Differential Evolution algorithm using bit operation based on swarm division.

In the proposed new algorithm, the location and velocity of each bat in space are represented through binary strings. We leverage logical operations to implement the mutation process where `+' represents `xor' operation and $\oplus$ represents `or' operation. $rand$ is a random number generated between 0 and 1. `$\cdot$' means only if $rand$$<$$F^t$ will the operations in the parentheses can be carried out. At iteration $t$ for the $i$th bat, if $rand$$<$$P^t$, we conduct the mutation on the velocity of the current bat as following, else we do nothing:

\begin{small}
\begin{equation}
\label{eq:entropye}
\begin{aligned}
&v_{ij}^{t}= x_{r5}^{t-1}+(x_{r1}^{t-1}\oplus x_{r2}^{t-1})\cdot F^{t}+(x_{r3}^{t-1}\oplus x_{r4}^{t-1})\cdot F^{t}
\end{aligned}
\end{equation}
\end{small}

Where $r_1$,$r_2$,$r_5$ are bats randomly selected in the same subgroup of the target while $r_3$,$r_4$ are bats selected in the distinctive subgroups.

$P^t$ is the mutation probability controlling whether bats perform the mutation operations or not. It varies adaptively with the number of iterations to obtain better search ability. $N_{T}$ is the largest number of iterations. $P^t$ is calculated as:

\begin{small}
\begin{equation}
\label{eq:entropye}
\begin{aligned}	
&P^{t}=\sqrt{\frac{t}{N_{t}-1}}
\end{aligned}
\end{equation}
\end{small}

At early iterations, each bat can make full use of its ability to search within a larger space with a small mutation probability. As the number of iterations increases, it is more likely for bats to mutate, which breaks the constraints of local minima for avoiding any premature convergence. 

$F^{t}$ is the shrinkage factor that is randomly generated between 0 and 1.The shrinkage factor $F$ regulates the variation of individuals by controlling the effect of the differential vectors. Larger values of $F$ is conductive to maintaining the diversity of the population while smaller values of $F$ enables bats with better local search ability. $F_{max}$ and $F_{min}$ are the maximum and the minimum of $F$, which is calculated as:

\begin{small}
\begin{equation}
\label{eq:entropye}
\begin{aligned}
&F^{t}=F_{min}+(F_{max}-F_{min})\cdot\frac{N_{t}-t}{N_{t}}
\end{aligned}
\end{equation}
\end{small} 

The modification of BA will ensure the algorithm with better exploration ability at early stage and higher exploitation at later stage. The division of the swarm enables efficient learning among similar individuals in the vicinity within each subgroup meanwhile sharing optimal information among the subgroups through the local minima. In this way, each bat can step to the global best gradually without being trapped into the local minim in avoid of its dramatical influence. By applying the Binary Differential Mutation, the proposed approach also increases the diversity of swarm and prevents from trapping into local minima, which also accelerates the rate of convergence. The pseudocode of the new algorithm is given in Algorithm 1. The relevant steps which are not introduced in our paper remains the same as original BA algorithm.
 
\begin{algorithm}[!htb]
	\caption{Improved Bat algorithm}   
	\label{alg:Framwork}   
	\begin{algorithmic}[1]  
		\REQUIRE ~~\\
		size of the swarm $N$; the number of subgroups $K$; maximum number of iterations $Max$; fitness function $f(x_i)$;
		\ENSURE  
		best solution of the swarm $G$;\\
		\STATE $//$ initialization\\
		\STATE generate initial population ($x_i$,$v_i$,$f_i$,$A_i$,$r_i$);\\ \STATE compute fitness function $f(x_i)$ 
		\STATE select global best solution $G_0$;\\ 
		\FOR{$t=1$ to $Max$}
		\STATE update controlling factors cf.eq.(3)(4)(6)(7);
		\STATE divide the swarm using K-means;\\
		\FOR{$n=1$ to $K$}
		\STATE select local minima at position $M_n$;\\
		\FOR{$i=1$ to round($N/k$)}
		\STATE retain previous best position $P_i$ 
		\STATE update the frequency $f_i$
		\IF{$i$ is non-local-minima bat} 
		\STATE update the velocity cf.eq.(1);\\
		\ELSE 
		\STATE update the velocity of local-minima cf.eq.(2);
		\ENDIF
		\STATE update the position $x_i$ 
		\IF{rand$>$$r_i$} 
		\STATE local search for the best solution $x_{new}$;\\
		\ENDIF
		\IF{rand$<$$A_i$ AND $f(x_i)$$<$$f(x_{new})$} 
		\STATE accept new solution;\\
		\STATE update $r_i$ and $A_i$;\\
		\ENDIF
		\IF{rand$<$$P^{t}$} 
		\STATE apply Binary Differential Evolution cf.eq.(5);\\
		\ENDIF
		\ENDFOR
		\ENDFOR
		\STATE determine best fitness $f(x_i)$ and update current $G_t$;
		\ENDFOR
		\RETURN $G_{Max}$  
	\end{algorithmic}  
\end{algorithm}

\section{Improved of Random Forest for flow classification}
\label{sec:algorithm2}
As an ensemble machine learning algorithm, RF \cite{Singh2014Big} has been widely used in processing high dimension dataset collaterally and preventing over-fitting to some extent. However, it degrades its performance in the minority classes of data when coping with an imbalanced dataset. Since the samples are randomly selected with replacement when building trees, the minority class with fewer samples is less likely to be selected and learned. In addition, we find that the cost of the minority class which is mis-classified is even higher than the majority ones which is urgent to improve its detection rate. Therefore, we need to optimize the algorithm as in the following ways.

\subsection{Weight initialization}

Usually, the weight of each sample is initialized similarly as $\frac{1}{N}$ ($N$ is the total number of the dataset) and the sum of them is 1. In this way, each sample is equal to be selected. In our mechanism, we initialize each training sample with a different weight according to the class that it belongs to. Corresponding to the original distribution of each class in our dataset, we initial the weights of five classes as 0.3, 0.15, 0.35, 0.05, 0.15. It reduces the weights of the majority class while boosting those of the minority class. In class j (j=1,2...5), the total number of samples is $N_{j}$ and the weight of each sample $w_{i}$ (i=1,2...$N_{j}$) is calculated as:

\begin{equation}
\label{eq:entropye}
\begin{aligned}
& w_{0,i}= \frac{w_{j}}{N_{j}}
\end{aligned}
\end{equation}

\begin{equation}
\label{eq:entropye}
\begin{aligned}
& W= (w_{0,1},w_{0,2}...w_{0,i}...w_{0,N})
\end{aligned}
\end{equation}

Where $w_{j}$ is the weight of class j. $N_{j}$ is the total number of samples in class j. $w_{i}$ represents the probability of each sample being selected from $N$ samples. In this way, the minority samples can be selected and paid more attentions instead of over-selecting the abundant samples in the majority class. We randomly select $N$ samples using the roulette wheel selection scheme from the original dataset with replacement to train each tree.

\subsection{Weight update}

After building each tree, we classsify the whole dataset and intend to update the weight of each sample according to the result it is classified. The weights of misclassified samples will be increased while samples correctly classified should be cut down. Consequently, we can stress more concern on the samples which are misclassified. Those samples with higher weights are more likely to be selected and learned in the next tree. We calculate the accuracy $a_m$ using the error rate $e_m$ of tree $m$ ($m$=1,2,...M) under the whole dataset and update the weight of the sample $w_{m+1,i}$ as following. We use $e^{a}$ to boost the weights of mis-classified samples and $e^{-a}$ for those correctly classified. M is the pre-defined number of trees for training. $Z_m$ is the scaling factor so that the total sum of weights remains 1.

\begin{equation}
\label{eq:entropye}
\begin{aligned}
& a_m = 0.5\cdot \ln \frac{1-e_m}{e_m}
\end{aligned}
\end{equation}
\begin{equation}
\label{eq:entropye}
\begin{aligned}
& w_{m+1,i} = \frac{w_{m,i}}{Z_{m}}\cdot \beta \cdot e^{\pm a}
\end{aligned}
\end{equation}
\begin{equation}
\label{eq:entropye}
\begin{aligned}
& Z_{m} = \sum_{i=1}^N w_{m,i}\cdot \beta \cdot e^{\pm a}
\end{aligned}
\end{equation}

We use a cost-sensitive way of learning by altering the weight of each sample. For the purpose of training the misclassified samples in distinctive classes specifically, we update the weight of each sample to various extent by controlling the $\beta$ factor in the Eq. (11) considering four situations in Table \uppercase\expandafter{\romannumeral1} \cite{Sun2010Improve}. $\beta$ is calculated as:
\begin{table}[htbp]
	\newcommand{\tabincell}[2]{\begin{tabular}{@{}#1@{}}#2\end{tabular}}
	\setlength{\abovecaptionskip}{0.cm}
	\setlength{\belowcaptionskip}{-20.cm}
	\footnotesize
	\centering
	\tabcolsep3pt
	\caption{$\beta$ calculated for different classes}
	\label{my-label}
	\begin{tabular}{|c|c|c|}
		\hline
		\tabincell{c}{Different classes} &\tabincell{c}{Samples \\ correctly classified} & \tabincell{c}{Samples \\ mis-classified}   \\ \hline
		The majority classes     & $2^{m-n}$    & $2^{n-m}$   \\ \hline
		The minority classes      & $2^{n-m}$         & $2^{m-n}$ \\ \hline
	\end{tabular}
\end{table}

In Table \uppercase\expandafter{\romannumeral1}, m and n are the sum of weights in the majority and minority classes. From the table we can see that, for the minority classes, we obviously increase the weights of misclassified samples, but slightly decrease those which are classified into the right class. For the majority classes, it is just the opposite. By this method, the minority samples can be selected and trained consistently and iteratively without degrading their weights remarkably when building trees. The weight update operation can also prevent samples from over-training repetitively.

\subsection{Weighted voting}

Now that the classification ability of each tree varies in different classes, the traditional method using the majority votes of all the trees for final result could not be used anymore. We introduce the weighted voting mechanism to the ensemble trees. Specifically, we compute the Accuracy Matrix of the classification accuracy $a_{m,j}$ of each tree $m$ in each class $j$ as in Table \uppercase\expandafter{\romannumeral2}. 

\begin{table}[htbp]
	\newcommand{\tabincell}[2]{\begin{tabular}{@{}#1@{}}#2\end{tabular}}
	\setlength{\abovecaptionskip}{0.cm}
	\setlength{\belowcaptionskip}{-20.cm}
	\footnotesize
	\centering
	\tabcolsep3pt
	\caption{Accuracy Matrix}
	\label{my-label}
	\begin{tabular}{|c|c|c|c|c|c|c|}
		\hline
		\tabincell{c}{Classes} &\tabincell{c}{Tree 1} & \tabincell{c}{Tree 2} & \tabincell{c}{...}  & \tabincell{c}{Tree m}  & \tabincell{c}{...} & \tabincell{c}{Tree M} \\ \hline
		Class0     & $a_{1,0}$    & $a_{2,0}$   & $...$ & $a_{m,0}$ & $...$ & $a_{M,0}$ \\ \hline
		Class1     & $a_{1,1}$    & $a_{2,1}$   & $...$ & $a_{m,1}$ & $...$ & $a_{M,1}$ \\ \hline
		Class2     & $a_{1,2}$    & $a_{2,2}$   & $...$ & $a_{m,2}$ & $...$ & $a_{M,0}$ \\ \hline
		Class3     & $a_{1,3}$    & $a_{2,3}$   & $...$ & $a_{m,3}$ & $...$ & $a_{M,3}$ \\ \hline
		Class4     & $a_{1,4}$    & $a_{2,4}$   & $...$ & $a_{m,4}$ & $...$ & $a_{M,4}$ \\ \hline

	\end{tabular}
\end{table}

The final result is computed according to Eq.(13). $f(x)$ is the ensemble result and $G_m(x)$ is the classification judgement of each tree $m$. The trees specializing in classifying different classes of samples can maximum its advantage in deciding the final result with higher weights. The process is illustrated in Figure 3. The colorful histograms represent accuracy of each tree in each class which is of different length. The figure shows classification of sample in class 2 as a example.

\begin{equation}
\label{eq:entropye}
\begin{aligned}
& f(x) = \sum_{m=1}^M a_{m,j}\cdot G_m(x)
\end{aligned}
\end{equation}

\begin{figure}[!t]
	\centering{\includegraphics[scale=0.25]{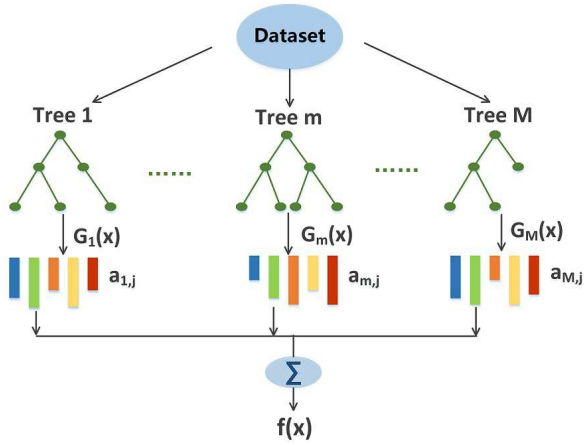}}
	\caption{Weighted voting}
\end{figure}

The modification of RF helps to strike a balance between over-learning in the majority class and directing more emphasis on the minority class, which enhances its performance in imbalanced dataset. Furthermore, the cost-sensitive learning of mis-classified samples also contributes to the overall accuracy. By using the weighted voting mechanism, various trees with distinguished abilities in classification can be strongly combined. The pseudocode of the algorithm is given in Algorithm 2.

\begin{algorithm}[!b]
	\caption{Improved Random forest}
	\begin{algorithmic}[1]
		\REQUIRE ~~\\
		the number of training data $N$ \\
		the number of trees $M$ \\
		samples selected for traning each tree $N_m$ \\
		the number of samples of $N_m$ $N_s$
		\ENSURE  
		the ensemble classification result $f(x)$ \\
		\STATE  $//$ initialization\\
		\STATE initialize weights of different classes $w_j$;
		\STATE compute weight of each sample $w_i$ in each class \\ cf.eq.(8);	
		\FOR{$m$=1 to $M$}
		\STATE training tree $m$ using sampling data $N_m$;
		\STATE calculate the  error rate $e_m$ and accuracy $a_m$ cf.eq.(9);
		\FOR{$i$=1 to $N_s$}
		\STATE update the weight of each sample $w_{m+1,i}$\\cf.eq.(10)(11)(12) and Table \uppercase\expandafter{\romannumeral1}; \\ 
		\ENDFOR
		\ENDFOR
		\STATE  determine the ensemble result $f(x)$ cf.eq.(13);
		\RETURN  $f(x)$ \\  
	\end{algorithmic}  
\end{algorithm}

\section{Evaluation result}
\label{sec:evaluation}
In this section, we conduct several numerical experiments to evaluate the proposed intrusion detection mechanisms.

\subsection{Dataset and evaluation metrics}
As the KDD Cup 1999 dataset \cite{Saxena2014Intrusion} has been widely used to evaluate various intrusion detection approaches, we perform a five-class flow classification using a subset of it after down-sampling in this paper. The distribution of both training and testing data marked by their attack type is summarized in Table \uppercase\expandafter{\romannumeral3}.

Generally, the performance of an intrusion detection system is evaluated in the light of precision (P), recall (R), F-score (F), accuracy (AC), and false alarm rate (FA). We desire a system with higher detection rate and lower false rate.  Another comparative metric $Cost$ is defined to measure the cost damage of misclassification for different attacks per sample. The relevant details of the dataset and evaluating metrics are introduced in our previous work \cite{Li2017A}.
\begin{table}[b]
	\setlength{\abovecaptionskip}{0.cm}
	\setlength{\belowcaptionskip}{-20.cm}
	\footnotesize
	\centering
	\caption{Distribution of data used in our evaluation}
	\label{my-label}
	\begin{tabular}{|c|c|c|c|c|}
		\hline
		\multirow{2}{*}{Class} & \multicolumn{2}{c|}{Training dataset} & \multicolumn{2}{c|}{Testing dataset} \\ \cline{2-5} 
		& No. of samples      & Percentage      & No. of samples      & Percentage     \\ \hline
		Normal                 & 17129               & 29.99\%         & 12183               & 32.52\%        \\ \hline
		Probe                  & 3107                & 5.44\%          & 1880               & 5.02\%         \\ \hline
		DoS                    & 35700               & 62.51\%         & 21705              & 57.94\%         \\ \hline
		U2R                    & 52                  & 0.09\%          & 228                 & 0.61\%         \\ \hline
		R2L                    & 1126                & 1.97\%          & 1468               & 3.92\%          \\ \hline
	\end{tabular}
\end{table}

\subsection{Performance Analysis}
We intend to evaluate the proposed mechanisms from three aspects. Firstly, we assess the optimality and convergence of the proposed BA algorithm for feature selection. Secondly, we estimate the detection ability of the enhanced RF algorithm for flow classification on the overall dataset as well as in different classes of flows respectively. Finally, we implement the combination of the above-mentioned algorithms in the proposed two-stage intrusion detection to make comparisons with the existing solutions. 

\subsubsection{Evaluations on the proposed Bat algorithm}\label{subsubsec5.3.1} 

There have been several improvements of BA to overcome its inherent shortcomings \cite{Enache2015Anomaly}\cite{Enache2015An}. We carry out a comparison with these methods as well as several typical Swarm Intelligence algorithms for feature selection together with RF for classification. The results are shown in Table \uppercase\expandafter{\romannumeral4} in terms of classification accuracy and false alarm rate. As noticed from the table, the proposed algorithm selects the features contributing more to differentiate attack traffic, which gives rise to higher accuracy and lower false rate. It is proved that the proposed algorithm for optimal feature selection performs well, which is conductive to achieve better performance in classification.

\begin{table}[t]
	\newcommand{\tabincell}[2]{\begin{tabular}{@{}#1@{}}#2\end{tabular}}
	\setlength{\abovecaptionskip}{0.cm}
	\setlength{\belowcaptionskip}{-20.cm}
	\footnotesize
	\centering
	\tabcolsep3pt
	\caption{Performance comparision using different algorithms}
	\label{my-label}
	\begin{tabular}{|c|c|c|}
		\hline
		\tabincell{c}{Different algorithms} & \tabincell{c}{Accuracy ($\%$)} & \tabincell{c}{FPR ($\%$)}  \\ \hline
		ACO                 & 94.25          & 4.78     \\ \hline
		PSO                 & 94.52          & 3.99     \\ \hline
		BA                 & 94.93          & 3.68     \\ \hline
		Reference[27]        & 95.35          & 2.74     \\ \hline
		Reference[28]        & 95.64          & 2.06     \\ \hline
		Our proposed algorthm        & 96.03          & 1.18     \\ \hline
	\end{tabular}
\end{table}

\begin{figure}[!b]
	\centering{\includegraphics[scale=0.1]{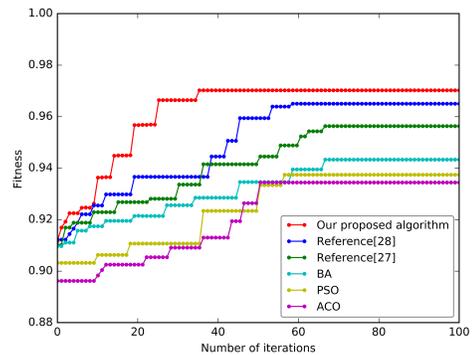}}
	\caption{Convergence comparision using different SI algorithms}
\end{figure}

In Figure 4, we verify the advantage of the proposed algorithm in finding a better subset of features with higher fitness within less iterations. It can be noticed that our algorithm converges faster than the others at about iteration 40 with a steeper slope, which lowers the time complexity. Furthermore, it obtains a higher fitness value after convergence and remains unchanged above the other curves. It is validated that the improvement of swarm division and the mutation mechanism help to escape from trapping into local minima and search for better solution. The linear time-varying parameters also strengthen individuals with dynamic searching ability to adjust different phases of iterations.

\begin{figure}[!t]
	\centering{\includegraphics[scale=0.15]{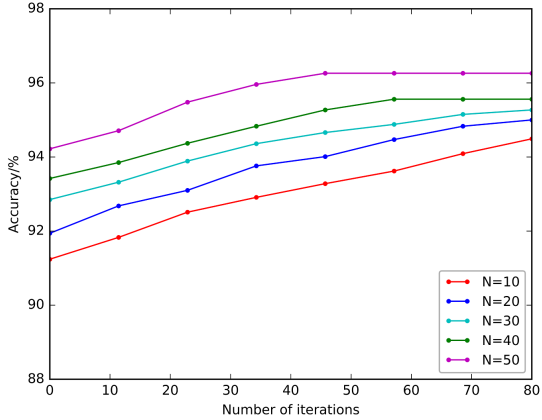}}
	\caption{Performance comparision using different numbers of individuals with various numbers of iterations}
\end{figure}

\begin{figure}[!b]
	\centering{\includegraphics[scale=0.12]{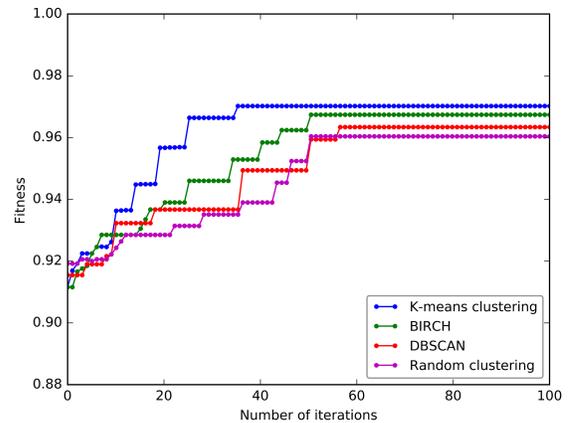}}
	\caption{Performance comparision using different methods of swarm division}
\end{figure}

Since the size of the swarm and the number of iterations are two critical parameters in process of solving optimization problems, we evaluate their influence with various values in Figure 5. It is illustrated that the accuracy of the algorithm enhances as the number of iterations increases with constant number of individuals in the swarm. It can be deduced that to some extent, the more iterations the swarm goes through, the better they evolves in finding the optimal solutions. Also, the algorithm will converge within restricted number of iterations in finding the approximate optimal solution. At the same iteration, we can see that a larger swarm with more individuals performs better than those smaller ones. It is because that a larger swarm with better diversity of population can communicate with each other more interactively without gathering close to the local minima, which points to a greater searching ability in larger area.

One of the most significant improvement of the proposed BA algorithm is the division of swarm into different populations. There are different kinds of methods to cluster individuals into populations. Some just randomly assign individuals into different clusters while the others use clustering algorithms in the light of various metrics. Therefore, we examine the influence of different clustering algorithms on the final BA algorithm and give the convergence results in Figure 6. It is apparent that using K-means \cite{Kanungo2002An} clustering algorithm achieves better performance with higher fitness and faster rate of convergence, since the adjacent individuals are clustered into same subgroup based on distances. On one hand, the whole swarm moves towards the current best position by learning within each subgroup as well as sharing knowledge between populations. On the other hand, each individual only learns partially from the local best in its subgroup and moves slightly in case of being badly affected by the local minima.

\begin{table}[!t]
	\newcommand{\tabincell}[2]{\begin{tabular}{@{}#1@{}}#2\end{tabular}}
	\setlength{\abovecaptionskip}{0.cm}
	\setlength{\belowcaptionskip}{-20.cm}
	\footnotesize
	\centering
	\tabcolsep3pt
	\caption{Performance comparision using different algorithms}
	\label{my-label}
	\begin{tabular}{|c|c|c|c|c|c|}
		\hline
		\tabincell{c}{Algorithms} & \tabincell{c}{Precision \\($\%$)} & \tabincell{c}{Recall \\($\%$)} & \tabincell{c}{F$\_$score \\($\%$)} &  \tabincell{c}{FPR \\($\%$)} & Cost    \\ \hline
		Decision Tree                 & 96.59          & 92.84
		& 95.42        & 4.79     & 0.2271            \\ \hline
		Adaboost                 & 97.42          & 93.21
		& 95.68       & 3.98     & 0.2032            \\ \hline
		RF                 & 98.09          & 93.84
		& 95.92       & 3.78     & 0.1738            \\ \hline
		SVM                 & 98.74          & 94.36
		& 96.55       & 2.75     & 0.1688            \\ \hline
		GBDT                 & 99.17          & 94.84
		& 96.72      & 1.68     & 0.1608            \\ \hline
		Proposed algorithm                & 99.51         & 95.17
		& 97.29        & 0.98     & 0.1302          \\ \hline
	\end{tabular}
\end{table}

\begin{table}[!t]
	\setlength{\abovecaptionskip}{0.cm}
	\setlength{\belowcaptionskip}{-20.cm}
	\footnotesize
	\centering
	\caption{Detection Performance comparision in different classes}
	\label{my-label}
	\begin{tabular}{|c|c|c|c|c|c|}
		\hline
		Algorithms & Normal & Probe & DoS & U2R & R2L \\ \hline
		Decision Tree & 95.63    & 95.47   & 99.02   & 6.25   & 8.27   \\ \hline
		Adaboost & 96.17      & 96.62     & 99.85   & 20.34   & 19.68   \\ \hline
		RF & 96.28      & 95.85     & 100   & 12.19   & 14.51   \\ \hline
		SVM & 97.72      & 97.55     & 99.78   & 12.49   & 12.72   \\ \hline
		GBDT & 98.54      & 98.79     & 100   & 27.65   & 21.45   \\ \hline
		Proposed algorithm  & 99.02   & 99.63 & 100   & 57.46   & 23.84  \\ \hline
	\end{tabular}
\end{table}

\subsubsection{Evaluations on the proposed Random forest algorithm}\label{subsubsec5.3.1} 

In Table \uppercase\expandafter{\romannumeral5}, we verify the advantage of improved RF algorithm through the performance of detection in contrast with the ordinary RF algorithm and other machine learning algorithms. All the  classification algorithms use features selected by the proposed BA as input. By altering the weights of samples, each tree in the forest can be trained more effectively through picking the samples which are more frequent to be mis-classified. By applying the weighted voting mechanism, the weight of each tree in a specific class is directly affected by its performance, which contributes distinctively to determining the final result. As we can see, it is obvious that the proposed method generates a better performance in every metric.

\begin{figure}[!b]
	\centering{\includegraphics[scale=0.17]{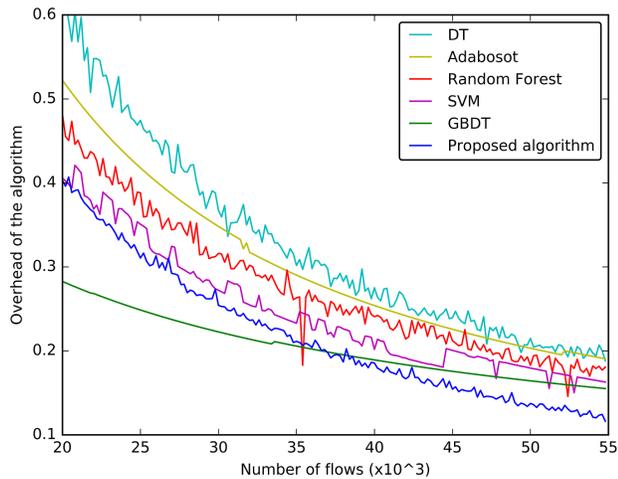}}
	\caption{Overhead produced by algorithms with different
		numbers of flows}
\end{figure}

We observe the detection performance of the improved algorithm in five classes individually comparing with the above-mentioned algorithms in Table \uppercase\expandafter{\romannumeral6}. In real network scenarios, some intrusions generate more connections than others which leads to an extreme unbalanced dataset for classification. The detection rate varies remarkably in different classes. Usually, those intrusions with fewer flows generate higher costs when mis-classified, so it is urgent to improve their performance. Thus, we solve the problem through altering the weights of samples accordingly to stress more attentions on those in the minority classes while avoiding over-fitting for the majority ones. It can be observed in Table \uppercase\expandafter{\romannumeral6}, it is apparent that the proposed RF algorithm improves the detection accuracy of minority intrusions dramatically while slightly increases the detection rate of the majority ones. The result indicates that the algorithm adaptively balances its training for samples in different classes and decides the final result according to its learning ability, which accelerates the performance in each class.  

\begin{table}[!t]
	\newcommand{\tabincell}[2]{\begin{tabular}{@{}#1@{}}#2\end{tabular}}
	\setlength{\abovecaptionskip}{0.cm}
	\setlength{\belowcaptionskip}{-20.cm}
	\centering
	\caption{Performance of different combinations of algorithms}
	\label{my-label}
	\begin{tabular}{|c|c|c|c|c|}
		\hline
		\multicolumn{2}{|c|}{\multirow{2}{*}{\tabincell{c}{Combination of \\algorithms}}} & \multirow{2}{*}{\tabincell{c}{No. of \\features}}  & \multirow{2}{*}{\tabincell{c}{Accuracy\\($\%$)}} & \multirow{2}{*}{\tabincell{c}{FPR\\($\%$)}} \\
		\multicolumn{2}{|c|}{}                                           &                                  &                                &                               \\ \hline
		Proposed BA  & Proposed RF  & 32  & 96.42  &0.98                     \\ \hline
		RF                               & GBDT                          & 22                               & 95.49      & 1.84                     \\ \hline
		Tree                             & SVM                            & 10                                & 94.41                         & 2.64                     \\ \hline
		Fisher                           & RF                            & 10     & 94.97                         & 2.35                     \\ \hline
		ReliefF                              & Adaboost                           & 8             & 95.32                         & 1.92                     \\ \hline
		IG                            & Decision Tree                            & 8       & 94.07                         & 4.35                     \\ \hline
		CFS   & LR                            & 18       & 93.22                         & 6.75                     \\ \hline				
	\end{tabular}
\end{table}

\begin{table}[!b]
	\newcommand{\tabincell}[2]{\begin{tabular}{@{}#1@{}}#2\end{tabular}}
	\setlength{\abovecaptionskip}{0.cm}
	\setlength{\belowcaptionskip}{-20.cm}
	\footnotesize
	\centering
	\tabcolsep3pt
	\caption{Performance comparision of different systems}
	\label{my-label}
	\begin{tabular}{|c|c|c|}
		\hline
		\tabincell{c}{Proposed system (and related works)}  & \tabincell{c}{Acurracy ($\%$)}  & \tabincell{c}{FPR ($\%$)}     \\ \hline
		Our system      & 96.42          & 0.98         \\ \hline
		Reference [26]    & 95.21          & 1.57          \\ \hline
		Reference [16]    & 94.56          & 1.83          \\ \hline
		Reference [30]      & 93.36          & 2.07     \\ \hline
		Reference [31]     & 92.42          & 2.82      \\ \hline
		Reference [19]      & 90.27          & 3.45     \\ \hline
	\end{tabular}
\end{table}

\begin{figure}[!b]
	\centering{\includegraphics[scale=0.2]{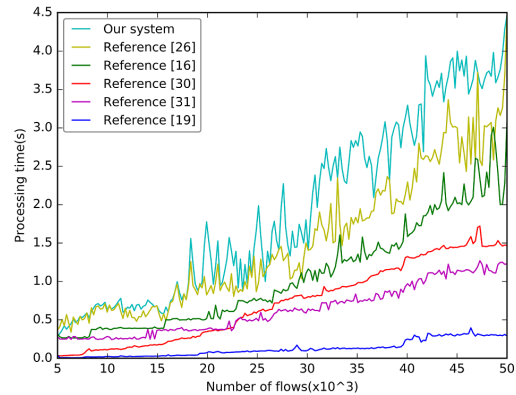}}
	\caption{Processing times of different systems}
\end{figure}

We measure the overhead that the proposed algorithm causes using the above-mentioned cost metrics. We compare the proposed algorithm with Random forest and the other machine learning algorithms in Figure 7. The metric indicates that the more mis-classified flows there are when making classification, the more overhead the algorithm generates for misclassification. As the number of flows grows, all of them become more well-training to make classification with less faults. It can be seen that our algorithm produces less overhead than its comparisons. 

\subsubsection{Evaluations on the performance of combined algorithms in the two stages}\label{subsubsec5.3.1}

We evaluate the combination of our improved algorithms comparing with several groups of traditional feature selection and machine learning algorithms in Table \uppercase\expandafter{\romannumeral7}. Since we know that the selection of algorithms for feature selection and traffic classification possesses a mutual influence on each other, we care more about the performance of the combination of them. As we can see, it is obvious that the proposed methods generate a better performance among all the combination
alternatives in every metric.

There have been several intelligent architectures proposed to detect and prevent network intrusions under SDN environment \cite{Silva2016ATLANTIC}\cite{Le2016Flexible}\cite{Li2017A}\cite{Ye2016An}\cite{Wang2017An}. We conduct a comparison with the previous results in terms of classification accuracy and error rate as described in Table \uppercase\expandafter{\romannumeral8}. The processing time of each approach using a portion of flows in the dataset is also illustrated in Figure 8 to evaluate the efficiency. It can be noticed that the proposed two-stage intrusion detection improves the classification accuracy with tolerable time consumption.

\section{Conclusion}
\label{sec:conclusion}
This paper presents an AI-based two-stage intrusion detection implemented in Software Defined IoT Networks. It leverages SDN contributing to status monitoring as well as traffic capturing under a global view. It integrates and coordinates two stages of IDS including feature selection and flow classification to detect novel intrusions with a self-learning ability. We improve Bat algorithm to select optimal features and design network flow classification methods by enhancing Random forest algorithm. Evaluation results validate the optimality of our proposed algorithms in achieving higher accuracy and lower overhead. The experiments also reveal that the system improves its detection ability without much time consumption compared with existing solutions.

In the future, we will implement this approach in a real network to traffic and evaluate the performance.

\bibliographystyle{IEEEtran}
\bibliography{IEEEfull,refer}

\end{document}